\documentclass[conference]{IEEEtran}
\IEEEoverridecommandlockouts
\usepackage{amsmath,amsfonts}
\usepackage{array}
\usepackage[ruled,vlined]{algorithm2e}
\usepackage{textcomp}
\usepackage{stfloats}
\usepackage{url}
\usepackage{verbatim}
\usepackage{graphicx}
\usepackage{cite}
\usepackage{algorithm2e}
\usepackage{svg}
\RequirePackage{letltxmacro}
\LetLtxMacro{\LaTeXtextbf}{\textbf}
\LetLtxMacro{\textbf}{\LaTeXtextbf}
\usepackage{cite}
\usepackage{amsmath,amssymb,amsfonts}
\usepackage{graphicx}
\usepackage{textcomp}
\usepackage{caption}
\usepackage{subcaption}
\usepackage[bookmarks=false]{hyperref} 
\usepackage{algpseudocode}
\usepackage{array}
\usepackage{enumitem,lipsum}
\setlist[itemize,enumerate]{leftmargin=*}
\usepackage{etoolbox}
\usepackage{xcolor}
\usepackage{graphicx}
\usepackage{mathptmx}

\usepackage{tcolorbox}
\tcbuselibrary{breakable}

\newcommand{\ie}{\textit{i.e.}}
\newcommand{\eg}{\textit{\eg,}}

\usepackage{tcolorbox}
\tcbuselibrary{breakable}

\usepackage{rotating}
\usepackage{multirow}
\usepackage[affil-it]{authblk} 

\ifCLASSINFOpdf

\else

\fi

\begin{document}

\title{AutoDFL: A Scalable and Automated Reputation-Aware Decentralized Federated Learning}


\author[1,2]{Meryem Malak Dif}
\author[1]{Mouhamed Amine Bouchiha}
\author[1]{Mourad Rabah}
\author[1]{Yacine Ghamri-Doudane}


\affil[1]{L3I, University of La Rochelle, La Rochelle, France}
\affil[2]{Ecole Nationale Supérieure d'Informatique, Algiers, Algeria}

\maketitle

\begin{abstract}
Blockchained federated learning (BFL) combines the concepts of federated learning and blockchain technology to enhance privacy, security, and transparency in collaborative machine learning models. However, implementing BFL frameworks poses challenges in terms of scalability and cost-effectiveness. Reputation-aware BFL poses even more challenges, as blockchain validators are tasked with processing federated learning transactions along with the transactions that evaluate FL tasks and aggregate reputations. This leads to faster blockchain congestion and performance degradation. To improve BFL efficiency while increasing scalability and reducing on-chain reputation management costs, this paper proposes AutoDFL, a scalable and automated reputation-aware decentralized federated learning framework. AutoDFL leverages zk-Rollups as a Layer-2 scaling solution to boost the performance while maintaining the same level of security as the underlying Layer-1 blockchain. Moreover, AutoDFL introduces an automated and fair reputation model designed to incentivize federated learning actors. We develop a proof of concept for our framework for an accurate evaluation. Tested with various custom workloads, AutoDFL reaches an average throughput of over 3000 TPS with a gas reduction of up to 20X.      

\end{abstract}


\IEEEpeerreviewmaketitle

\begin{tcolorbox}[breakable,boxrule=1pt,colframe=black,colback=white]
\scriptsize Paper accepted at IEEE/IFIP Network Operations and Management Symposium (NOMS'2025) IEEE, 2025.
\end{tcolorbox}

\section{Introduction}
\IEEEPARstart{F}{ederated} learning (FL) has emerged as a promising decentralized training paradigm, enabling devices to collaboratively train a model while keeping their training data local \cite{flsrv1,noms2024data}. In this paradigm, a central server aggregates the global model based on local models trained on each device’s raw data. However, traditional centralized FL systems face significant security challenges. The reliance on a central server creates a single point of failure, making the system vulnerable to a variety of attacks including model poisoning and inference attacks \cite{poisoningnoms2022,inferenceDSC2023}. Additionally, the accuracy of the global model can be compromised by malicious devices that upload low-quality model weights. Furthermore, the lack of transparency complicates the verification of the integrity of individual contributions \cite{srvfl2024}. 

\quad To address these vulnerabilities, decentralized federated learning (DFL) has been proposed as an effective solution \cite{srvdfl2024}. In DFL, there is no central server to coordinate or aggregate the model updates. Instead, the participating devices (clients) communicate directly with each other in a peer-to-peer (P2P) network. Devices exchange model updates among themselves, and each participant can independently aggregate the updates it receives from its peers. One way to decentralize federated learning is via blockchain technology \cite{srvbfl2023}. The central server is replaced with a decentralized peer-to-peer infrastructure, running a consensus mechanism to ensure the reliability and efficiency of the FL processes. Blockchain provides transparent and tamper resistance mechanisms that record a complete history of FL transactions, allowing for the tracking and auditing of any malicious behavior. This transparency fosters trust among participants in the federated learning process, as all actions—such as model updates and data access—are verified and recorded as part of the protocol. 

\quad Despite its benefits, DFL, particularly blockchain-based federated learning (BFL) faces many challenges. Due to the transparent nature of blockchain, BFL frameworks are vulnerable to free-riding attacks, where malicious or dishonest trainers (clients) deliberately avoid contributing useful work while still benefiting from the final aggregated model \cite{dflfreeriderinfocom2024,srvdfl2024}. Furthermore, challenges remain regarding the computational expense and scalability of BFL systems, particularly as the number of participants in the network increases, especially in cross-device scenarios. Although some approaches seek to enhance training and aggregation efficiency\cite{nomsDFL2023,noms2024data}, they introduce high computational complexity, which may hinder practical deployment in blockchain infrastructure. More robust algorithms and frameworks are necessary to manage large-scale federated learning effectively\cite{srvbfl2023}, as existing BFL systems face efficiency and scalability issues that must be addressed to ensure their viability in real-world applications\cite{valente2023federated}. In addition, ensuring the security and trustworthiness of the federated learning process is crucial. To foster active participation and cooperation among nodes in the network, it is essential to incorporate incentive management mechanisms within the BFL system. However, while these mechanisms are vital for regulating participants' behavior and maximizing system efficiency, implementing them on-chain for enhanced transparency introduces additional complexity and overhead to the underlying blockchain \cite{rollupthecrowd}. This, in turn, can impact the overall performance and scalability of the system. Building an efficient and trustworthy blockchain-based federated learning (BFL) system comes at a cost, requiring a careful balance between incentivization management and the operational demands of the blockchain network. \\

Motivated by the above challenges, our contribution presented in this paper covers the following points:
\begin{itemize}
    \item A blockchain-powered, fully decentralized platform to manage federated learning (AutoDFL), leveraging zk-Rollups (Layer-2) solutions for enhanced scalability by reducing the burden on the main chain (Layer-1).
\item An automated, transparent, and fair reputation model designed for federated learning scenarios, robust against common malicious behaviors and adversarial actions, independent of task publishers' personal assessments or utility evaluation functions, emphasizing resilience against collusion, bad-mouthing, and other security attacks.
\item Secure and robust federated learning smart contract automation utilizing a Decentralized Oracle Network (DON).
\item A comprehensive proof of concept implemented using emerging technologies, and the code is publicly available on GitHub\footnote{\href{https://github.com/meryemmalakdif/AutoDFL}{https://github.com/meryemmalakdif/AutoDFL}}. Extensive experimental evaluations to validate the efficiency and scalability of the proposed framework.
\end{itemize}

\section{Related work} \label{sec:relatedWork}

\quad This section presents an analysis of existing literature on decentralized blockchain-based federated learning, examining current studies from three key perspectives: efficiency, automation, and scalability.

\quad Several frameworks have been proposed to enhance the efficiency of decentralized federated learning by improving its fairness and trustworthiness. For instance, the authors in \cite{chen2024credible} introduced a credible and fair framework that employs smart contract-enabled computation to evaluate client contributions, addressing issues such as model poisoning and free-rider behavior. Similarly, \cite{yuan2024trustworthy} proposed a trustworthy federated learning scheme (TWFL) that utilizes two-trapdoor homomorphic encryption to secure gradients and combat malicious actions in Web 3.0, enhancing resilience against inference and poisoning attacks. In terms of incentivizing participation, \cite{xu2021besifl} proposed a fully decentralized BFL paradigm that mitigates malicious nodes through an accuracy-driven detection approach, while employing a contribution-based incentive mechanism with a token reward scheme to encourage credible nodes to participate in the learning process. Similarly, FGFL \cite{gao2022fgfl} emphasizes reliable incentive management by incorporating auditing and reputation-based server selection, which ensures transparency in reward distribution. TrustRe, described in \cite{zhang2021blockchain}, extends this concept by introducing a reputation-based worker selection mechanism that utilizes blockchain as a transparent ledger for recording worker reputations. This framework employs Proof of Reputation (PoR) for model aggregation, where reputable workers are selected as bookkeepers for the blockchain ledger to enhance the quality of the global model.
Moreover, a BFL framework introduced in \cite{9997114} employs a novel consensus mechanism called Proof of Interpretation and Selection (PoIS), which aggregates feature attributions to distinguish significant contributors from outliers while utilizing a credit function for fair reward distribution. Despite the effort made in these studies and others\cite{qi2022high,mugunthan2022blockflow,ruckel2022fairness}, the cost of implementing their proposed incentive mechanism in cross-device BFL networks remains a concern. In other words, achieving fairness and trust among BFL actors comes at a price, which is the impact of implementing such mechanisms on the underlying blockchain.

The scalability and cost-effectiveness of blockchain-based federated learning (BFL) systems are crucial for large-scale deployments involving thousands of devices. The above BFL approaches face the issues of high latency and storage, as well as limited throughput. This is mainly due to large model parameters and the intensive computations conducted on the blockchain, which may hinder their real-world applicability and broader adoption. To address the storage bottleneck, several BFL systems utilize the InterPlanetary File System (IPFS) \cite{benet2014ipfs}, storing data off-chain in a distributed file system with content addresses serving as unique pointers to each file across devices. Other frameworks focus on layer 1 (L1) scalability improvements like sharding, where the network is split into smaller shards to process transactions in parallel \cite{liu2023survey}.
The research in \cite{moudoud2023multi} introduces a blockchain-based multitask federated learning framework designed for the Metaverse, enabling concurrent model learning and utilizing blockchain sharding to enhance system throughput while prioritizing devices with reliable behavior and valuable datasets, thereby optimizing bandwidth and securing interactions to reduce vulnerabilities to malicious activities. ScaleSFL, \cite{madill2022scalesfl}, also employs a sharded architecture where each shard functions as an independent chain, with nodes collaboratively training local models and offloading raw weights to IPFS while retaining only their corresponding hashes on the shard ledger. 
However, dividing the network into smaller, independent shards introduces vulnerabilities, such as single-shard takeovers \cite{han2022analysing}. 

Although many efforts have been made to improve BFL scalability, layer 2 (L2) solutions that leverage off-chain mechanisms for more efficient transactions and computations have not been adopted yet. Additionally, none of the above studies have attempted to automate the aggregation and the evaluation of trainers' contributions while reducing the on-chain computational cost by offloading the computation to an Oracle network.
This gap in current research needs to be investigated to determine whether these solutions can effectively address the scalability and cost-effectiveness challenges within BFL frameworks.

\section{Proposed Framework} \label{sec:proposedframework}

\subsection{System Model} 
\quad In AutoDFL, we identify five key roles contributing to its functionality and integrity: validators, task requesters, training agents, contribution evaluators, and aggregators. \textit{Validators} $\{V_i\}$ are responsible for ensuring the correctness of transactions and verifying that they adhere to established protocols. Once transactions are validated, they participate in Quorum-based Byzantine Fault Tolerance \cite{yin2018hotstuff} consensus to agree on the next block in a permissioned blockchain. \textit{Task publishers} $\{TP_i\}$ (or requesters) initiate the training process by defining and posting specific tasks. \textit{Training agents} $\{TA_i\}$ subscribe to tasks that align with their expertise and utilize their local data to train models and gain rewards. \textit{Training evaluators} $\{TE_i\}$ assess the utility of each training agent’s contributions, while \textit{aggregators} $\{AG_i\}$ combine local model weights from multiple trainers to create a cohesive global model.
\subsection{Threats Model} 

\quad In AutoDFL, it is assumed that all entities, including $TPs$, $TAs$, $TEs$, $AGs$, and $Vs$ can exhibit malicious behavior. Moreover, all participants involved in the system are assumed to have no inherent trust in one another and are primarily motivated by self-interest, which may lead them to engage in harmful activities or collude in coordinated attacks. Additionally, the framework operates under the assumption that no central trusted authority exists and that at least $2/3$ of the blockchain and DON nodes behave correctly. Given these assumptions, AutoDFL is designed to address a broad spectrum of potential attacks, as discussed in the following:

\begin{itemize}

\item \textbf{Sybil attacks.}
A Sybil attack occurs when an attacker creates multiple fake identities to exploit the reputation system. 
    
\item \textbf{False-reporting attacks.}
A false-reporting attack occurs when a task publisher denies the high quality of a model produced by an honest trainer participating in a task. This denial can prevent the trainer from being rewarded for their contributions and efforts, allowing malicious requesters to obtain high-quality products without compensating the associated fees.


\item \textbf{Free-riding attacks.}
A free-riding attack occurs when a trainer subscribes to a training task but submits an arbitrary model instead of making a genuine effort to train, or intentionally skips training rounds out of laziness. Such behavior aims to gain a portion of the reward without providing equivalent contributions or support. This malicious activity can degrade the global model's convergence rate and slow down the overall training process.


\item \textbf{Collusion attacks.}
A collusion attack occurs when trainers collaborate with requesters to lower a target trainer’s reputation or inflate their own. 

\item \textbf{Inference attacks.} A major threat to data privacy, as the data characteristics of participating trainers can be inferred from their submitted model weights. 

\item \textbf{Bad-mouthing attacks.}
Bad-mouthing occurs when the evaluator or the aggregator submits an incorrect evaluation or an inaccurate aggregated global model, respectively. 


\item \textbf{Whitewashing attacks.}
A whitewashing attack occurs when a dishonest worker re-enters the system with a new identity to reset their low reputation score. 

\end{itemize}

\subsection{System Architecture}

\quad The layered architecture of AutoDFL is depicted in Fig. \ref{fig:arch}, which includes the following layers:

\begin{figure}[t]
    \centering
    \includegraphics[width=\columnwidth]{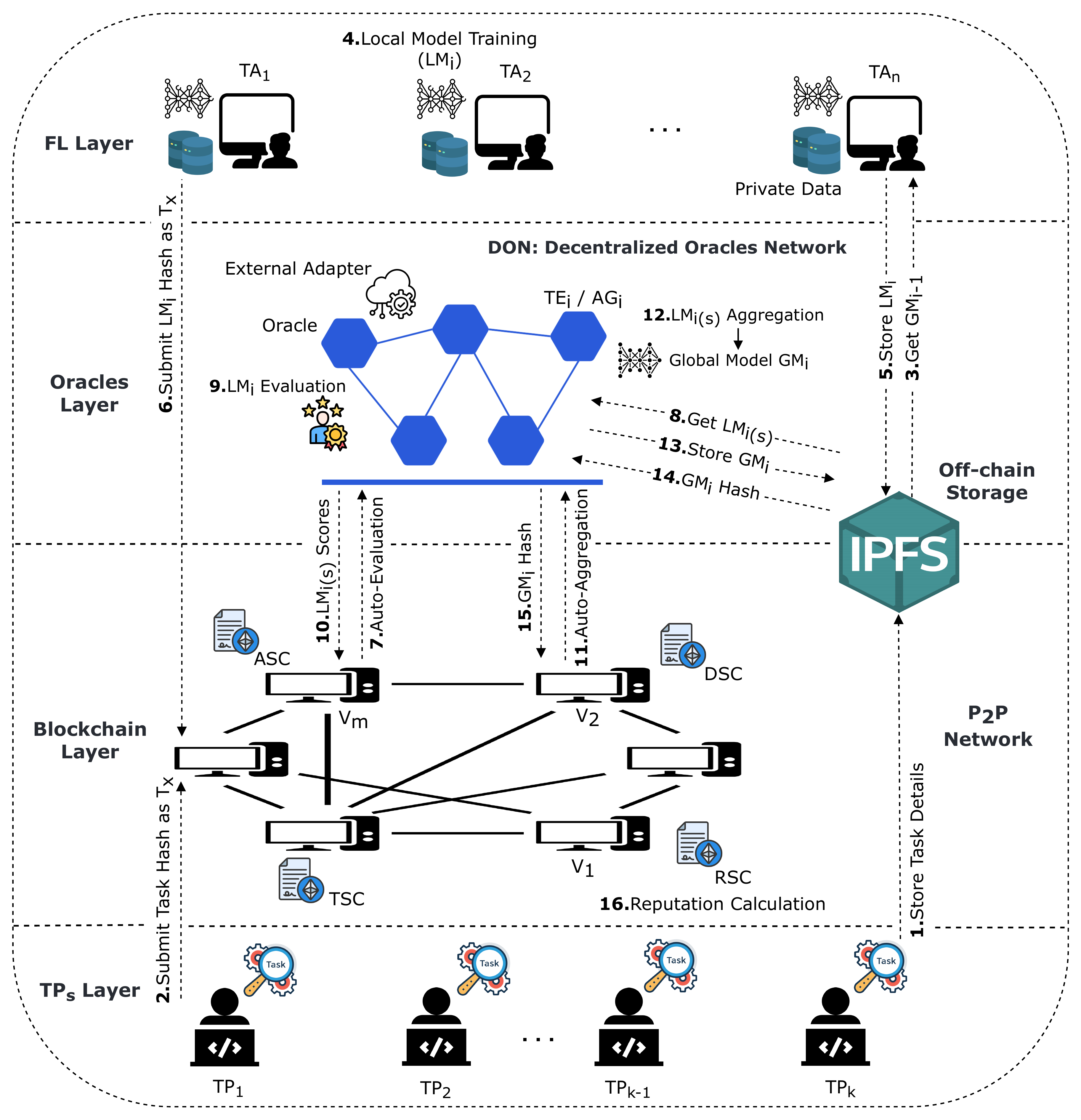} 
        \caption{System Architecture. ASC refers to the Access Smart Contract, which enforces role-based control and regulates permissions, granting access solely to authorized users.}
    \label{fig:arch}
\end{figure}

\subsubsection{Task Publishers Layer} Entities interested in creating and disseminating tasks initiate the process by allowing clients to discover and subscribe to tasks relevant to their interests.
\subsubsection{Federated Learning Layer} This layer enables decentralized model training while preserving data privacy. Trainers, who are clients participating in specific tasks, use their local data to train models and submit only the updated model weights to IPFS and then to the blockchain. This approach allows for the aggregation of knowledge from diverse data sources while ensuring that sensitive information remains on client devices.
\subsubsection{Dual-Layered Blockchain with zk-Rollups} 
\begin{figure}[t]
    \centering
    \includegraphics[width=\linewidth]{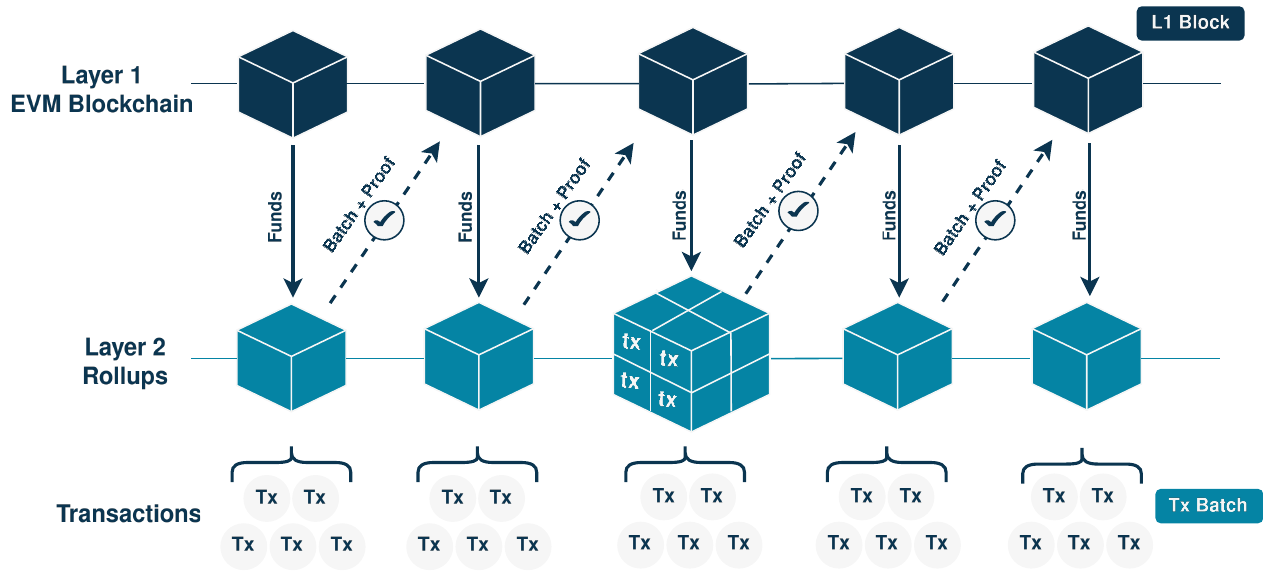}
    \caption{Dual-Layered Blockchain Design: L1 is an Ethereum Virtual Machine (EVM)-based blockchain, L2 is powered by zk-Rollups.}
    \label{fig:rollups}
\end{figure}

The system implements a dual-layer blockchain architecture, utilizing zk-rollups to enhance scalability and efficiency, as illustrated in Fig. \ref{fig:rollups}. The base layer operates as a permissioned blockchain network, relying on a Quorum-based Byzantine Fault Tolerance consensus mechanism to maintain a secure and consistent ledger. To minimize the computational and storage demands on this base layer, the second layer employs zk-rollups \cite{thibault2022blockchain} \footnote{\href{https://blog.matter-labs.io/introducing-zk-sync-the-missing-link-to-mass-adoption-of-ethereum-14c9cea83f58}{https://blog.matter-labs.io/introducing-zkSync}}. In this architecture, zk-rollups bundle multiple transactions off-chain, drastically reducing the workload on the main blockchain and improving transaction throughput. This process involves executing batches of transactions off-chain and generating validity proofs using zero-knowledge proofs, ensuring the correctness of the state changes. Once the validity proof is generated, it is submitted to the base layer along with a concise summary of the state changes, allowing the base layer to verify the proof and confirm that the transactions were processed correctly off-chain. This approach eliminates the need to post each transaction individually on-chain, reducing fees while guaranteeing the entire process's integrity and security. This dual-layered model enables the system to efficiently manage large volumes of transactions while preserving the core attributes of blockchain technology: security, decentralization, and transparency.

\subsubsection{Off-Chain Storage} We use InterPlanetary File System (IPFS), a decentralized peer-to-peer file storage network, to manage the off-chain storage of model weights and task details. By offloading substantial data to IPFS, the system reduces the storage burden on the blockchain, resulting in faster transaction times and significantly lower costs. This integration addresses the scalability, speed, and cost challenges often faced in traditional blockchain-based federated learning systems, ultimately enhancing the performance and feasibility of large-scale deployments.
\subsubsection{Decentralized Oracles Network (DON)} The integration of the InterPlanetary File System (IPFS) necessitates a robust mechanism for data synchronization and retrieval. Since blockchain networks cannot directly access external data sources \ie, smart contracts cannot interact with data not stored on-chain—such as model weights in our framework, decentralized oracles act as intermediaries that facilitate secure and efficient data transfer from IPFS to on-chain smart contracts. Therefore, their implementation in AutoDFL enables a safe data flow between off-chain and on-chain components.
\subsection{System Workflow}

\quad The workflow of AutoDFL involves the following steps:
\begin{enumerate}    
\item \textbf{Publish task.}
A task requester publishes a training task (steps 1-2 in Fig.\ref{fig:arch}), detailing the task description, model architecture, required accuracy, and a validation dataset for evaluating local models. This is done by invoking the $publishTask$ function (Algo. \ref{algo1new}) within the tasks smart contract (TSC). The task details are stored off-chain on IPFS, and the resulting cryptographic hash is recorded on the blockchain for transparency and reference. To ensure fair compensation, the requester deposits a reward for trainers, which is securely locked on the blockchain using a deposit/escrow smart contract (DSC). This ensures transparency and accountability in the reward distribution.

\begin{algorithm}[th]
\footnotesize
\caption{Publish Task} \label{algo1new}
\begin{algorithmic}[1] 
    \State\KwIn{ taskID, modelCID, descriptionCID }

    \State \textbf{begin}
    \State \ \ \ \texttt{Assert}(isTaskPublisher(msg.sender) = true)
           \State \quad tasks.push(Task(\{
                \State \quad \quad taskId: taskID,
                \State \quad \quad modelCID: modelCID,
                \State \quad \quad description: descriptionCID,
                \State \quad \quad publisher: msg.sender,
                \State \quad \quad trainers: new address[](0),
                \State \quad \quad currentRound: 0,
                \State \quad \quad state: "selection"
            \}))

    \State \textbf{end}
    \end{algorithmic}
\end{algorithm}

\item \textbf{Select trainers.}
Trainer selection is conducted on-chain using a function in TSC that evaluates the reputation scores of registered clients linked to the published task. These reputation scores are securely stored on the blockchain. The contract selects the most reliable clients based on their high ratings and records their account addresses in the task details on the blockchain.
\item \textbf{Train and submit model.} (steps 3-6 in Fig.\ref{fig:arch}) A selected trainer locks collateral into DSC to guarantee their commitment and deter malicious behavior. The trainer then retrieves the model architecture and other relevant information stored on IPFS using the task hash stored on the blockchain. With this information, the trainer trains a machine learning model using their local data. To enhance data privacy and protect sensitive information, differential privacy is applied by adding carefully calibrated noise to the model's weights ($\mathit{w}' = \mathit{w} + \mathit{n}$) \cite{xiong2020comprehensive}. This approach aims to make it difficult to infer or extract specific knowledge from the model weights while ensuring that the introduction of noise does not significantly compromise accuracy. The trainer uploads the updated model weights ${w}'$ to IPFS, and the generated hash is recorded on the blockchain by invoking the $submitLocalModel$ function within TSC (Algo.\ref{algo2}). 

\begin{algorithm}[th]
\footnotesize
\caption{Submit Local Model} \label{algo2}
\begin{algorithmic}[1]
    \State\KwIn{ taskID, round, localModelCID}

    \State \textbf{begin}
    \State \ \ \ \texttt{Assert}(isTrainerInTask(msg.sender,taskID) = true)
    \State \ \ \ models[taskID][round][msg.sender] $\gets$ localModelCID

    \State \ \ \ modelsSubmitted[taskID][round][msg.sender] $\gets$ true

    \State \textbf{end}
\end{algorithmic}
\end{algorithm}

\item \textbf{Evaluate contributions.} (steps 7-10 in Fig.\ref{fig:arch}) To evaluate trainers' contributions, a training evaluator $TE$ implemented on the Oracle network retrieves local model weights from IPFS using their hashes stored on the blockchain. $TE$ assesses each trainer's performance by measuring their model's accuracy against the validation dataset provided by the task publisher $TP$. Based on this evaluation, the oracle assigns a score to each trainer reflecting their model's quality. After aggregating scores from multiple oracles, the average final scores for the round are calculated and securely stored on-chain.

\item \textbf{Aggregate models.} (steps 11-15 in Fig.\ref{fig:arch})
Upon completion of the evaluation, oracles initiate the aggregation of local models by combining their weights. To improve the efficiency of our aggregation protocol, we adapt the traditional FedAvg aggregation algorithm \cite{li2019convergence} by taking into account the scores of $TAs$. Each local model's weight $w_i$ is multiplied by the trainer's previously calculated score $s_i$, which reflects its utility or contribution. 
\begin{equation}
    w_g = \frac{\sum_{i=1}^{n} s_i \cdot w_i}{\sum_{i=1}^{n} s_i}
\end{equation}

where, \( w_g \) refers to the global model weights, \( s_i \) is the score of the \( i \)-th trainer, \( w_i \) are the weights of the \( i \)-th local model, and \( n \) is the total number of trainers.

\item \textbf{Calculte new reputation.} (step 16 in Fig.\ref{fig:arch})
Upon completing a training task, oracles invoke the \textit{calculateNewRep} function to refresh the reputation scores stored on-chain using the proposed reputation model. 
Details on the reputation update are presented in the following section.
\end{enumerate}

\section{Reputation Model}
\label{sec:reputationmodel}
\quad After describing the architecture of AutoDFL and its
main components, we will now delve into the mathematical
details of the proposed reputation model.
\subsection{Objective Reputation}
\quad In federated learning (FL), maintaining a reliable assessment of participants' performance is crucial. To guarantee this, we develop a deterministic objective evaluation. More precisely, we introduce a local objective reputation to evaluate the interaction between $TP$ and $TAs$, focusing solely on their performance in the current task upon its completion. $TAs$ who achieve high accuracy and consistently demonstrate positive behavior are assigned elevated scores, while those exhibiting laziness or malicious intent receive lower scores. 
\begin{equation}
\label{equObjRep}
O_{{\mathit{rep}}_\mathit{i}} = \mathit{scoreAuto} \cdot \frac{\nu_c}{\nu_t} \cdot \left(1 - \max\left(\frac{\mathit{ND}_i - \tau}{1 - \tau}, 0\right)\right)
\end{equation}
\begin{equation}
\mathit{ND}_i = \frac{\mathit{D}_i}{\max(D_1, D_2, \ldots, D_n)}
\label{normalization}
\end{equation}
\begin{equation}
\mathit{D}_i = \sqrt{\sum_{j=1}^{m} (w_j^{\mathit{LM}} - w_j^{\mathit{GM}})^2}
\label{euclidean}
\end{equation}

\quad The computation of a trainer $i$'s objective reputation, as represented by the formula in Equ. \ref{equObjRep}, considers the performance score, denoted as $scoreAuto$, which reflects the quality of the trainer $i$'s contribution during the task as measured by the DON; the task completeness level, measured by the ratio of completed rounds $\nu_{c}$ to the total rounds $\nu_{t}$, which assesses the participation rate to indicate how many training rounds each trainer actively engaged in, as some trainers may deliberately skip rounds; and a penalty component based on the Euclidean distance $D_{i}$ between the trainer $i$'s local model weights $w^{\mathit{LM}}$ and the global model parameters $w^{\mathit{GM}}$ in the last round (Equ. \ref{euclidean}). The Euclidean distance $D_{i}$, determined by the DON, is scaled relative to the maximum distance observed across all trainers (Equ. \ref{normalization}), producing a normalized distance $ND_i$. When $ND_i$ falls below a specified threshold $\tau$, no penalty is applied. However, if $ND_i$ exceeds $\tau$, a sanction is imposed that increases in proportion to the distance. The threshold $\tau$ is adjustable and can be set as the average of distances among all trainers to ensure fair penalization.



\subsection{Subjective Reputation}
\quad Subjective reputation allows task publishers to assess the reliability of trainers based on historical interactions, representing a trust value that reflects the task publisher’s direct opinion of the trainer. To achieve this, we adapt the subjective trust logic~\cite{subtl}, where each interaction between a training agent $TA$ and a task publisher $TP$ results in an opinion, denoted as \( O_{TP \to TA} = (b, d, u) \), being formed to express $TP$’s belief in the trustworthiness of $TA$. In this representation, $b$, $d$, and $u$ correspond to belief, disbelief, and uncertainty, respectively, with $b$, $d$, $u \in [0, 1]$ and $b$ + $d$ + $u = 1$.

\begin{equation}
\begin{cases}
    b = (1 - u) \frac{\alpha}{\alpha+\beta} \\
    d = (1 - u) \frac{\beta}{\alpha+\beta} \\
    u = 1 - I_f \\
    I_f = \frac{X_{TA \to TP}}{X_{TP}}
\end{cases}
\end{equation}

\quad The interaction frequency $I_{f}$ reflects how often $TP$ has interacted with $TA$, measured by the proportion of $TA$'s interactions with $TP$ ($X_{TA \to TP}$) relative to $TP$'s total interactions with all trainers in the system ($X_{TP}$). A higher interaction frequency reduces uncertainty, indicating greater confidence and familiarity between $TP$ and $TA$.

\quad The parameters $\alpha$ and $\beta$ represent the number of tasks in which the trainer’s performance was judged as good or poor, respectively. These are calculated as:
\begin{equation}
\begin{cases}
    \alpha = \sum_{j=1}^{n} \theta \cdot \mathit{C}_{j} & \text{Good performance in task $j$} \\
    \beta = \sum_{j=1}^{n} (1-\theta) \cdot \mathit{C}_{j} & \text{Otherwise}
\end{cases}
\end{equation}
where $C_{j}$ represents the recency of task $j$, giving more weight to recent tasks, and $\theta$ determines the impact of good or poor behavior. To discourage malicious actions and encourage truthful contributions, the model intentionally assigns a higher weight to poor interactions than to good ones. 

The subjective reputation score is then computed as:
\begin{equation}
\mathit{S}_{\text{rep}} = b + \sigma \cdot u
\end{equation}
where $\sigma$ is a configurable parameter reflecting the uncertainty weight.
\subsection{Local Reputation}
\label{sec:localrep}
\quad Local reputation reflects the trustworthiness of a trainer, factoring in their performance in both the current task and previous ones. It is calculated using the following formula:
\begin{equation}
\mathit{L}_{\text{rep}} = \mathit{O}_{\text{rep}} \cdot \gamma + \mathit{S}_{\text{rep}} \cdot (1 - \gamma)
\label{equ:localrep}
\end{equation}
where $\gamma$ is a configurable parameter that controls the importance of the objective reputation versus the subjective reputation. Putting more weight on $O_{rep}$ means that performance on the current task is more important than historical behavior.

\subsection{Reputation Update}
\quad In AutoDFL, each new participant is assigned an initial reputation value $R_{init}$. This value can be derived from the average reputation scores of all existing trainers or can be a fixed value determined by the consortium's operating system.
We believe that a trainer can only improve their reputation through good performance and effective investment in their data and system capabilities to train FL models. The reputation management smart contract (RSC) is triggered at the end of each task to update the overall reputation scores of the participating trainers. Initially, the local reputation $L_{rep}$ is computed, as described by Equ.\ref{equ:localrep}, and this value is subsequently used to update the overall reputation scores of the participants involved in the task, along with their current reputation.

\begin{equation} \label{equRep}
\mathit{R}_{\text{i}} =
\begin{cases}
\omega \cdot \mathit{R}_{\text{i-1}} + (1 - \omega) \cdot \mathit{L}_{\text{rep}}  & \mathit{L}_{\text{rep}} \geq \mathit{R}_{\text{min}} \\
(1 - \omega) \cdot \mathit{R}_{\text{i-1}} + \omega \cdot \mathit{L}_{\text{rep}} & \mathit{L}_{\text{rep}} < \mathit{R}_{\text{min}}
\end{cases}
\end{equation}

\begin{equation} \label{equWeight}
\omega = \tanh_{\lambda}(\mathit{N}) = \frac{1 - e^{-\lambda \mathit{N}}}{1 + e^{-\lambda \mathit{N}}}
\end{equation}

Where $R_{i-1}$ represents the current reputation value prior to the most recent task, and $R_{i}$ denotes the new reputation score. $N$ refers to the number of tasks a trainer has engaged in since joining the network while $\omega$ is a weighting factor derived from a hyperbolic tangent function that normalizes $N$, rather than a static factor with $\omega \in [0, 1]$. This allows longer-tenured active participants to enhance their reputation while ensuring that mistakes are not overly tolerated. Equ.\ref{equWeight} ensures that as trainers engage in more tasks, their potential for reputation enhancement increases, provided they continue to adhere to appropriate behavior. Note that, $\lambda$ is a parameter that determines the rate of increase for $\omega$, with its specific value established by the consortium's operating system.

The formula defined by Equ.\ref{equRep} adapts the update weights based on whether the trainer exhibits good or bad behavior. We define a threshold $R_{min}$, which represents the critical line of trust (\ie, any participant with a reputation below this threshold is deemed untrusted).
If a trainer's local reputation $L_{rep}$ is above the threshold, their new reputation is computed with a focus on prioritizing their previous reputation. This approach assumes that trainers are generally expected to exhibit good behavior, allowing their reputation to increase gradually. Consequently, a trainer must perform well on a regular basis to attain a high reputation. This way AutoDFL incentivizes trainers to maintain consistent performance. Conversely, if a trainer's local reputation $L_{rep}$ falls below the threshold, their new reputation is calculated with a greater emphasis on the local reputation. This adjustment amplifies the impact of poor behavior on the trainer’s overall reputation. 

\section{Security Analysis}

\quad In this section, we examine the key security risks and the measures implemented by AutoDFL to counter them.

\begin{itemize}
\item \textbf{Sybil attacks.} Our system mitigates this by enforcing strict identity verification, with designated administrators (\ie, consortium members) exclusively authorized to add or remove users under consortium oversight. This is done through on-chain permissioning with majority voting.

\item \textbf{False-reporting attack.} AutoDFL addresses this threat through two key mechanisms. First, we do not grant task requesters the privilege of assessing a trainer's performance utility. This ensures that task requesters cannot falsely claim low performance for a trainer who has performed well. Second, prior to the commencement of training, the task requester is required to lock the payment as a deposit in a smart contract. This deposit serves as a commitment to the system, preventing the task requester from repudiating the payment to a deserving trainer.

\item \textbf{Free-riding attack.} We mitigate this threat by employing a robust evaluation mechanism that rigorously assesses the utility of each trainer's contribution prior to reward distribution. Rewards are allocated exclusively to those who meet high-quality standards, ensuring that only genuine and meaningful contributions receive recognition.

\item \textbf{Collusion attacks.} In AutoDFL, reputation growth is based solely on a trainer's utility during tasks, and task requesters do not have the privilege of assessing trainer performance. This prevents requesters from falsely reporting low performance for well-performing trainers and from boosting the reputation of trainers with whom they are colluding.

\item \textbf{Inference attacks.} To mitigate inference attacks, our system employs differential privacy by adding noise to the model weights, ensuring that model accuracy remains largely intact while effectively obscuring individual data characteristics, thus enhancing privacy for trainers. By adding a suitable noise, $TAs$ can make it statistically infeasible for attackers to succeed. 

\item \textbf{Bad-mouthing attacks.} AutoDFL addresses bad-mouthing the same way it prevents bad-collusion by preventing $TPs$ from evaluating $TAs$. Additionally, by leveraging DON, the ratings of individual models and their aggregated outputs are generated and cross-verified by multiple independent nodes. This approach guarantees a robust evaluation and aggregation process by validating data accuracy and correctness before its integration into the blockchain.

\item \textbf{Whitewashing attacks.} A whitewashing attack occurs when a dishonest worker uses a new identity to re-enter the system and reset their low reputation score. Our system mitigates this by permitting rejoining only with approval from the consortium through majority voting.

\end{itemize}

\section{Evaluation and Results} \label{sec:evaluation} 
\quad To evaluate the AutoDFL framework, we developed a proof of concept. The code with all the technical details is available on GitHub \footnote{\href{https://github.com/meryemmalakdif/AutoDFL}{https://github.com/meryemmalakdif/AutoDFL}}.
\subsection{Environment} 
\quad We used our platform dedicated to evaluating blockchain solutions. It consists of two HPE ProLiant XL225n Gen10 Plus servers. Each server is equipped with two AMD EPYC 7713 64-core processors and 512 GB of RAM (2x256 GB).
\subsection{Experimental Setup}

\quad We developed our smart contracts in Solidity\footnote{\href{https://docs.soliditylang.org/en/v0.8.27/}{https://docs.soliditylang.org}} and deployed them on the Ethereum blockchain, utilizing the Geth\footnote{\href{https://geth.ethereum.org/}{https://geth.ethereum.org/}} client for network operations and Web3.py\footnote{\href{https://web3py.readthedocs.io/en/stable/}{https://web3py.readthedocs.io}} for smooth interactions between the client application and the smart contracts. To facilitate oracle integration, we incorporated Chainlink\footnote{\href{https://chain.link/}{https://chain.link/}} nodes enhanced with customized external adapters built using a Python API. For L2 scaling, we leveraged zkSync\footnote{\href{https://zksync.io/}{https://zksync.io/}} Rollups to address the scalability limitations of the main Ethereum network. To rigorously assess the efficiency and scalability of our proposed solution, we conducted extensive testing and benchmarking using the Hyperledger Caliper\footnote{\href{https://github.com/hyperledger/caliper-benchmarks}{https://github.com/hyperledger/caliper-benchmarks}} framework, which allows us to evaluate blockchain latency and throughput. Furthermore, we analyzed additional performance indicators, including gas fees derived from the logs generated by the running nodes.
For federated learning, we utilized the MNIST dataset of handwritten digits \cite{LeCun2005TheMD}, shuffling the dataset and distributing portions of the training data among clients while employing the LeNet-5 model\cite{tan2019improved} to enhance image recognition for this specific application.

\begin{figure}[t]
    \centering
  \includegraphics[width=0.95\columnwidth]{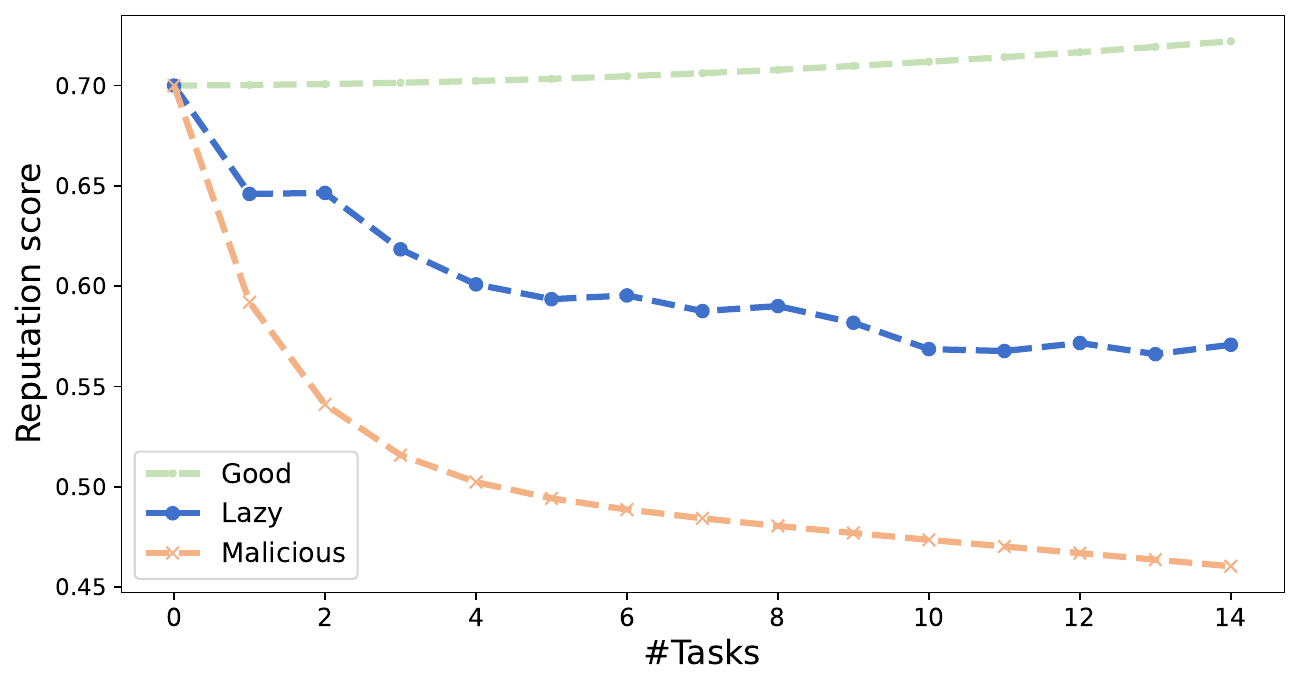}
    \caption{Reputation Dynamics of Profiles With Varying Behaviors: Good, Malicious,
and Lazy Participation.}
    \label{fig:repchanges}
\end{figure}


\subsection{Reputation Effectiveness} 
\quad We begin the experimental study by evaluating the effectiveness of the proposed reputation model. We simulate the behavior of three different $TAs$:
\begin{itemize}
    \item \textbf{Good Behavior:} This profile consistently engages in training models with genuine effort.

    \item \textbf{Malicious Behavior:} This profile attempts to deceive the system by submitting random model weights without performing any actual training. 

   \item \textbf{Lazy/Inconsistent Behavior:} This profile intermittently skips rounds, actively participating in only a portion of them, typically missing between 40\% to 60\% of the rounds.

\end{itemize}

Changes in reputation evolution of the three $TAs$ profiles over a sequence of tasks is illustrated in Fig. \ref{fig:repchanges}. Good behavior leads to a gradual but steady increase in reputation. The system rewards continuous, honest contributions, and the small but consistent rise in reputation highlights the importance of sustained participation. Malicious and lazy behaviors on the other hand lead to significant declines in the reputation of the trainers. The reputation score of a malicious $TA$ drops sharply, demonstrating the system's ability to detect and penalize dishonest actions. Additionally, the model acknowledges occasional contributions from lazy trainers, as the rate of reputation change is linked to the number of missed rounds.

\subsection{Blockchain Performance Evaluation}

\quad Our experimental evaluation focuses on three key metrics:
\begin{itemize}
    \item \textbf{Throughput:} The number of successful transactions per second (TPS).
\item \textbf{Latency:} The time delay (s) between the initiation of a transaction and its confirmation.
\item \textbf{Gas:} A unit measuring the computational work required to perform operations, influenced by the complexity of tasks, computational steps, and the amount of data processed.
\end{itemize}
\quad The results presented below pertain to the evaluation of a comprehensive federated learning scenario, encompassing task creation and reputation updating, with specific emphasis on the performance of four key functions: \textit{publishTask, submitLocalModel, calculateObjectiveRep}, and \textit{calculateSubjectiveRep}.\\

\begin{figure}[t]
    \centering
  \includegraphics[width=\columnwidth]{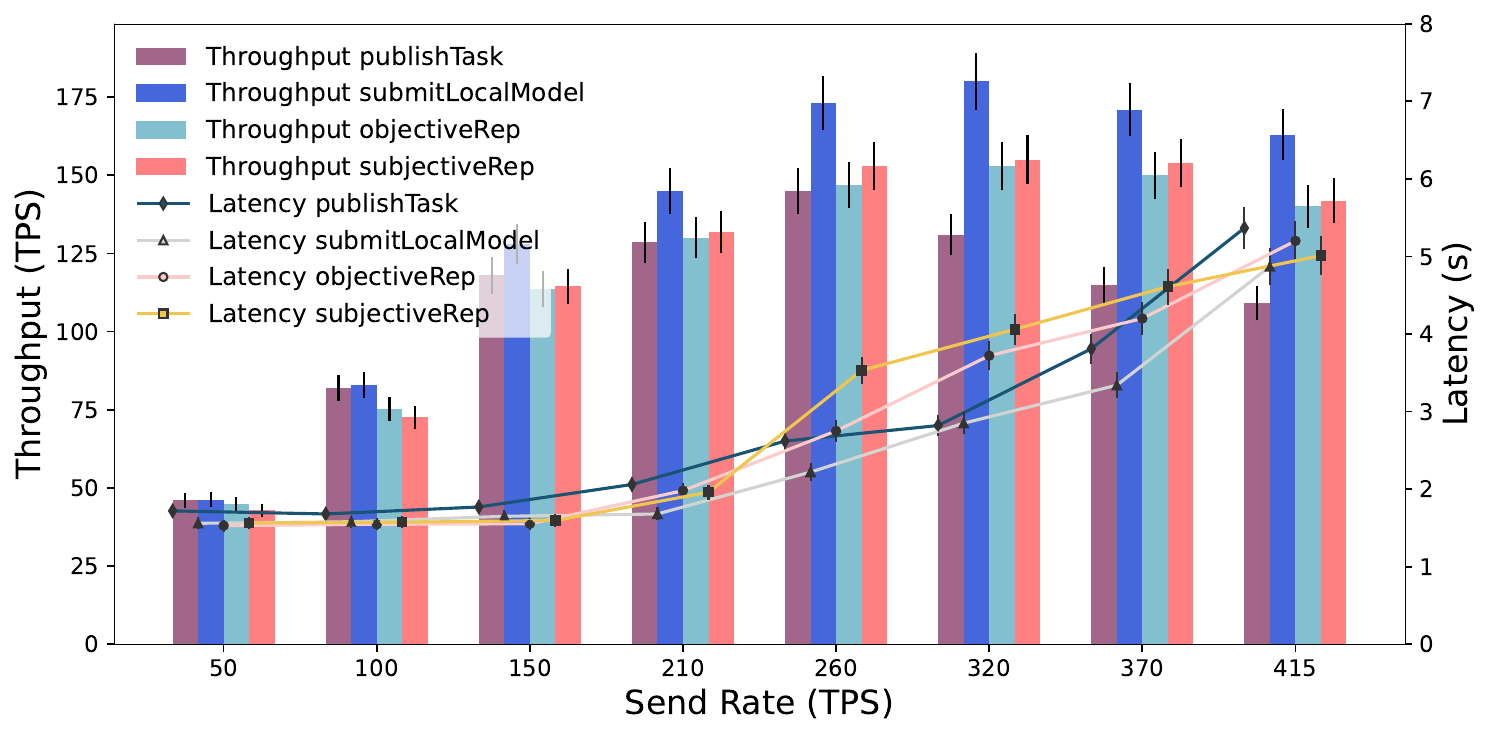}
    \caption{L1 Throughput and latency comparison under different
workload types.}
    \label{fig:throu_latency}
\end{figure}

\textbf{1. L1 Throughput and Latency.}
The throughput and latency values for each function under different send rates are illustrated in Fig. \ref{fig:throu_latency}. Initially, the pattern is evident: throughput and latency both increase as the transaction send rate rises. The lightest function, $submitLocalModel$, achieves the highest peak throughput, reaching around 180 (TPS), with a send rate of 320 (TPS). This performance surpasses that of the other functions, primarily due to the on-chain computational overhead and additional storage requirements associated with those functions.
As the send rate increases, each function experiences a steady rise in throughput and a corresponding increase in latency, reflecting the system's growing load. However, once the send rate surpasses a certain threshold, the L1 blockchain begins showing signs of congestion. Throughput declines, while latency continues to rise sharply, indicating that the network can no longer process transactions as efficiently. This behavior underscores L1's limitations.

\begin{figure}[t]
    \centering
  \includegraphics[width=0.95\columnwidth]{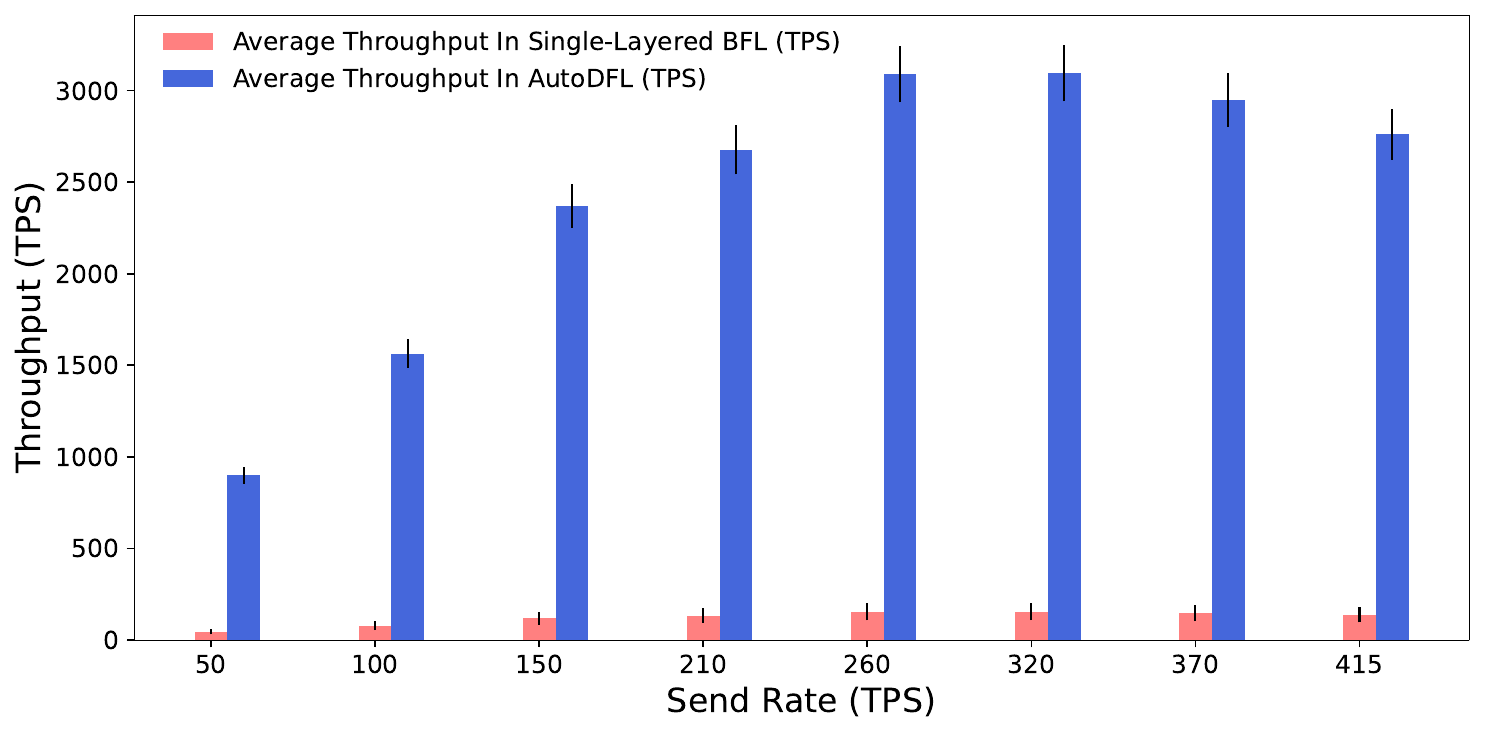}
    \caption{Average Throughput Comparison: Single-Layered BFL vs AutoDFL.}
    \label{fig:l1vsl2throuphput}
\end{figure}

\textbf{2. L2 vs L1 Performance.} We measure the gas cost for multiple simultaneous function calls under different workloads in both traditional single-layer (L1) and dual-layer (L2) blockchains. In the dual-layer implemented using zk-rollup, the validation of L2 batches on L1 involves three distinct phases: \texttt{commit}, \texttt{prove}, and \texttt{execute}. During these stages, batches are sequentially committed, proven, and executed on L1, with each step incurring its own gas consumption. The overall gas cost is the cumulative sum of the gas expended in these three phases.
The results, as demonstrated in Tab. \ref{gascost}, highlight the dual-layer (L2) approach efficiency in terms of gas consumption across all functions when compared to the single-layer (L1) approach. This reduction can be attributed to the key feature of zk-rollups: they shift much of the computation off-chain while preserving security and data availability on-chain through the use of zero-knowledge proofs.\\
On L1, gas costs increase linearly with the number of function calls, as each call incurs a fixed gas cost. This results in the total gas expenditure being the product of the gas cost per function and the number of calls made. In contrast, within the L2 (zk-rollup) environment, the gas costs associated with the \texttt{prove} and \texttt{execute} phases remain relatively stable, even as the number of transaction calls increases. This stability is due to zk-rollups' efficiency in managing proof verification and block execution, regardless of the transaction volume.
However, gas consumption during the \texttt{commit} phase shows variability depending on the complexity of the function. For example, the $publishTask$ function, which stores a substantial amount of information, incurs the highest gas costs due to the increased computational effort required for processing and storing the additional data. \\
Despite the variation in the \texttt{commit} phase, the overall gas consumption in zk-rollups remains lower than that in the traditional single-layer blockchain. This efficiency underlines the benefits of adopting L2 solutions to reduce gas costs by up to 20 times. The key factor contributing to this reduction is the batching employed by zk-rollups. For function calls up to 20, only a single batch is committed, allowing for the aggregation of up to 20 transactions into one batch. When the number of function calls exceeds 20, a new batch is committed and submitted to L1.
To calculate L2 throughput, the batch size is multiplied by the L1 throughput. For example, with a batch size of 20 transactions and a L1 throughput of 150 (TPS), AutoDFL can achieve a throughput of $20 \times 150 = 3000 \text{ (TPS)}$. Fig. \ref{fig:l1vsl2throuphput} shows how zk rollups amplify throughput to handle a significantly higher volume of transactions compared to L1. Finally, Tab. \ref{overhead}, presents the average end-to-end L2 latency of multiple calls for the four main functions in AutoDFL, demonstrating that processing multiple transactions simultaneously takes only a few seconds.


\begin{table}[t]
\centering
\caption{Gas consumption of the four main functions in AutoDFL: L1 vs L2}
\setlength{\tabcolsep}{3pt} 
\renewcommand{\arraystretch}{1.2} 
\resizebox{0.5\textwidth}{!}{ 
\begin{tabular}{|c|c|c|c|c|c|c|c|}
\hline
\multirow{3}{*}{\textbf{Function}} & \multirow{3}{*}{\textbf{\#Calls}} & \multicolumn{5}{c|}{\textbf{Dual Layer (L2)}} & \multirow{3}{*}{\textbf{Single Layer (L1)}} \\ \cline{3-7}
      &  & \multicolumn{2}{c|}{Commit} & \multirow{2}{*}{Verify} & \multirow{2}{*}{Execute} & \multirow{2}{*}{Total} &  \\ \cline{3-4}

      &  & \#Batches & Consumed Gas & &  &  &  \\
      \hline 

\multirow{4}{*}{ \begin{turn}{-30} \textit{publishTask} \end{turn}}      
   &  5     & 1 & 61300  & 27272        & 23964   & \textbf{112536}  & \textbf{910931} \\
 & 20    & 1 & 127052  & 29892        & 26964   & \textbf{183908} & \textbf{3566355} \\
  & 50    & 3 & 359896  & 29904        & 26584   & \textbf{416384} & \textbf{8878594} \\ 
  & 100   & 5 & 685639  & 29904        & 26572   & \textbf{742115} & \textbf{17736655}  \\ \hline
      \multirow{4}{*}{ \begin{turn}{-30} \textit{submitLocalModel} \end{turn}}      
    &  5     & 1 & 44588  & 27272        & 23964   & \textbf{95824}  & \textbf{251108} \\ 
& 20    & 1 & 67112  & 29880        & 26560   & \textbf{123552} & \textbf{930181} \\ 
& 50    & 3 & 185092  & 29892        & 26584   & \textbf{241568} & \textbf{2288330} \\ 
& 100   & 5 & 354956  & 27284        & 26584   & \textbf{408824} & \textbf{4135650}  \\ \hline


       \multirow{4}{*}{ \begin{turn}{-30} \textit{calculateObjectiveRep} \end{turn}}      

      & 5     & 1 & 37662  & 27272        & 23952   & \textbf{88886}  & \textbf{265815} \\ 
& 20    & 1 & 41164  & 29904        & 26608   & \textbf{97676} & \textbf{983156} \\ 
 & 50    & 3 & 125884  & 29892        & 26584   & \textbf{182360} & \textbf{2205124} \\ 
 & 100   & 5 & 216688  & 29940        & 26584   & \textbf{273212} & \textbf{4299248}  \\ \hline


     \multirow{4}{*}{ \begin{turn}{-30} \textit{calculateSubjectiveRep} \end{turn}}      & 5     & 1 & 36020  & 27284        & 23976   & \textbf{87280}  & \textbf{196296} \\ 
    &  20    & 1 & 36532  & 29904        & 26608   & \textbf{93044} & \textbf{715350} \\ 
      & 50    & 3 & 109180  & 29940        & 26608   & \textbf{165728} & \textbf{1760587} \\ 
      & 100   & 5 & 181544  & 29892        & 26584   & \textbf{238020} & \textbf{3523732}  \\ \hline
           
\end{tabular}
}
\label{gascost}
\end{table}

\begin{table}[t]
\centering
\caption{Time overhead (s) for different functions}
\setlength{\tabcolsep}{4pt} 
\renewcommand{\arraystretch}{1.2} 
\resizebox{0.4\textwidth}{!}{
\begin{tabular}{|c|c|c|c|c|c|c|}
\hline
\textbf{\#Calls} & \textbf{1} & \textbf{5} & \textbf{10} & \textbf{20} & \textbf{50} & \textbf{100} \\ \hline
\textit{publishTask}            & 1.145 & 1.564 & 2.452 & 3.201 & 7.514 & 14.785 \\ \hline
\textit{submitLocalModel}      & 0.176 & 0.731 & 1.285 & 2.297 & 6.524 & 14.280 \\ \hline
\textit{calcObjectiveRep}      & 0.214 & 0.686 & 1.304 & 2.627 & 6.756 & 14.660 \\ \hline
\textit{calcSubjectiveRep}     & 0.221 & 1.037 & 1.495 & 3.784 & 8.726 & 17.075 \\ \hline
\end{tabular}
}
\label{overhead}
\end{table}


\section{Conclusion} \label{sec:conclusion}
\quad In this paper, we presented AutoDFL, a scalable and automated reputation-aware decentralized federated learning framework. AutoDFL employs zk-Rollups as a Layer-2 scaling solution, thereby enhancing performance while maintaining the same level of security as the underlying Layer-1 blockchain. Furthermore, AutoDFL incorporates an automated and impartial reputation model, designed to encourage FL agents to behave consistently and correctly. Finally, to provide a more realistic evaluation, we developed a complete proof of concept for our framework, using emerging tools such as zkSync rollups and Chainlik Oracles. Evaluated with different customized workloads, AutoDFL reaches an average throughput of over 3000 TPS with a gas reduction of up to 20X.  

\quad For future work, cryptographic techniques such as zero-knowledge proofs (ZKP), to enable secure verification of training and aggregation without disclosing sensitive data, and homomorphic encryption (HE), to perform computations (training and aggregation) on encrypted data, can be introduced to further enhance the privacy and overall safety of our DFL framework.



\bibliographystyle{IEEEtran}
\bibliography{ref}

\end{document}